\documentclass[preprint]{elsarticle}
\usepackage[nodots]{numcompress}

\usepackage{lineno}
\journal{TBC}

\begin{document}
%TC:ignore
\begin{frontmatter}



\title{MRI-guided cardiac radiotherapy using a 1.5 T MR-linac: surveying emerging patterns of care}

\author[umcu]{O. Akdag} \author[umcu,UMCU_CIG]{S. Mandija} \author[umcu]{J. Pomp} \author[umcu]{M.P.W. Intven} \author[MDA]{X. Chen} \author[MDA]{J. Yang} \author[AHN]{S. Oh} \author[AHN]{M. Trombetta} \author[GC_Perth,FionaStanley]{H. Tan} \author[FionaStanley]{J. De Leon} \author[FionaStanley,UNSW,UniWollongong]{M. Jameson} \author[BaskentAnkara]{E. Efe} \author[BaskentAnkara,BaskentAdana]{C. {\"O}nal} \author[umcu]{M.F. Fast}
\affiliation[umcu]{organization={Department of Radiotherapy, University Medical Center Utrecht},
            addressline={Heidelberglaan 100},
            city={Utrecht},
            postcode={3584 CX},
            state={Utrecht},
            country={The Netherlands}}

\affiliation[UMCU_CIG]{organization={Computational Imaging Group for MR Diagnostics and Therapy, Center for Image Sciences, University Medical Center Utrecht},
            addressline={Heidelberglaan 100},
            city={Utrecht},
            postcode={3584 CX},
            state={Utrecht},
            country={The Netherlands}}

\affiliation[MDA]{organization={Department of Radiation Physics, University of Texas MD Anderson Cancer Center},
            addressline={1515 Holcombe Blvd},
            city={Houston},
            postcode={TX 77030},
            state={Texas},
            country={United States of America}}

\affiliation[AHN]{organization={Department of Radiation Oncology, Allegheny Health Network},
            addressline={501 Penn Ave},
            city={Pittsburgh},
            postcode={PA 15222},
            state={Pennsylvania},
            country={United States of America}}

\affiliation[GC_Perth]{organization={GenesisCare, Oncology},
            addressline={24 Salvado Rd},
            city={Perth},
            postcode={Wembley WA 6014},
            state={West-Australia},
            country={Australia}}

\affiliation[FionaStanley]{organization={Department of Radiation Oncology, Fiona Stanley Hospital},
            addressline={11 Robin Warren Dr},
            city={Perth},
            postcode={Murdoch WA 6150},
            state={West-Australia},
            country={Australia}}

\affiliation[UNSW]{organization={Department of Clinical Medicine, UNSW Sydney},
            addressline={High St},
            city={Sydney},
            postcode={Kensington NSW 2052},
            state={New South Wales},
            country={Australia}}

\affiliation[UniWollongong]{organization={Department of Medical and Radiation Physics, University of Wollongong},
            addressline={Northfields Ave},
            city={Wollongong},
            postcode={Wollongong NSW 2522},
            state={New South Wales},
            country={Australia}}

\affiliation[BaskentAnkara]{organization={Department of Radiation Oncology, Başkent University Faculty of Medicine},
            % city={Çankaya/Ankara},
            postcode={06490},
            state={Ankara},
            country={T{\"u}rkiye}}

\affiliation[BaskentAdana]{organization={Department of Radiation Oncology, Başkent University Faculty of Medicine},
            addressline={Adana Dr. Turgut Noyan Research and Treatment Center},
            % city={Adana},
            postcode={01120},
            state={Adana},
            country={T{\"u}rkiye}}

%% Abstract
\begin{abstract}
% \textit{Background \& purpose}: The occurrence of cardiac tumours is rare. Cardiac lesions are typically treated by surgical resections, often combined with chemotherapy to attain increased overall survival. Surgical procedures for cardiac tumours are complex and invasive, with significant carry procedural risks. MRI-guided radiotherapy (MRgRT) is a noninvasive treatment option facilitating conformal dose deliveries through a combination of soft-tissue imaging and adaptive radiotherapy techniques, including cardiorespiratory motion management. To assess the application of MRgRT for cardiac tumours, we conducted a patterns of care analysis with users of the 1.5T Unity MR-linac.

\textit{Background:}  Cardiac tumours are rare and treated by surgical resections, which are complex, invasive and carry procedural risks. MRI-guided radiotherapy (MRgRT) noninvasively facilitates conformal dose deliveries using soft-tissue imaging and adaptive radiotherapy techniques.
%Cardiac tumors are rare and typically treated by surgical resections, often combined with chemotherapy to attain increased overall survival. Surgical procedures for cardiac tumours are complex and invasive, with significant procedural risks. MRI-guided radiotherapy (MRgRT) is a noninvasive treatment option facilitating conformal dose deliveries through a combination of soft-tissue imaging and adaptive radiotherapy techniques, including cardiorespiratory motion management. 

\noindent \textit{Aim:} To assess the application of MRgRT for cardiac tumours, we conducted a patterns-of-care analysis with users of the 1.5T MR-linac.
%To assess the application of MRgRT for cardiac tumours, we conducted a patterns of care analysis with users of the 1.5T Unity MR-linac.

\noindent \textit{Materials \& methods}:  A survey was distributed to users of the 1.5T MR-linac that treated patients with cardiac tumours. The survey included 30 questions concerning the patient cohort, imaging, treatment planning/simulation, radiotherapy treatment and treatment outcome.
%A survey was developed with 30 questions and distributed to users of the 1.5T Unity MR-linac (Elekta AB, Stockholm, Sweden) that treated patients with cardiac tumours. The survey included questions concerning the patient cohort, imaging modalities, treatment planning/simulation, radiotherapy treatment and treatment outcome.

\noindent \textit{Results}: Users from six international institutes completed the survey reporting twelve cardiac MRgRT treatments between 2021-2024. The median age[range] of the patients was 59[16--81] years with 50\% of the cases concerning the treatment of primary tumours. The prescribed dose ranged between 30--60 Gy, with 30 Gy being prescribed the most (59\%). In 75\% of the cases, the treatment plans were delivered in five fractions with 6--8 Gy per fraction. Daily images were acquired with T1- and T2-weighted MRI and respiratory motion monitoring was performed in 92\% of the cases using cine imaging. A single case was treated with intrafraction motion management using a novel vendor-provided gating solution on the MR-linac. Treatment outcomes were reported for 50\% of the cases. All, but one case, attained local control using MRgRT without serious adverse events (grade$\geq$3).
% Users from six international institutes completed the survey reporting twelve cardiac MRgRT treatments on the Unity MR-linac between 2021-2024. The median age[range] of the patients was 59[16--81] years with 50\% of the cases concerning the treatment of primary tumours and 50\% concerning the treatment of metastases. The prescribed dose ranged between 30--60 Gy, with 30 Gy being prescribed the most (59\%). In 75\% of the cases, the treatment plans were delivered in five fractions with 6--8 Gy per fraction. Daily images were acquired with T1- and T2-weighted MR imaging and respiratory motion monitoring was performed in 92\% of the cases using cine imaging. A single case was treated with intrafraction motion management using a novel vendor-provided gating solution on the Unity MR-linac. Treatment outcomes were reported for 50\% of the cases. All, but one case, attained local control using MRgRT and no serious adverse events (grade$\geq$3) were reported.

\noindent \textit{Conclusion}: This study provides real-world insights into the feasibility and early outcomes to aid the development of cardiac MRgRT treatment recommendations and protocol harmonization.
%This study provides real-world insights into the feasibility and early outcomes to aid the development of cardiac MRgRT treatment recommendations and protocol harmonization.
% Emerging patterns of care indicate that MRgRT is technically feasible for cardiac tumours. The findings of this study could aid the development of treatment recommendations and harmonize cardiac MRgRT treatment protocols.
\end{abstract}

%% Keywords

\end{frontmatter}

%%\linenumbers

%TC:endignore
\section{Introduction}\label{sec:Introduction}
The occurrence of cardiac tumours is rare (frequency range of 0.001\%--0.3\% based on post-mortem analysis) and sarcomas account for most primary malignancies\cite{yanagawa_surgery_2018}. Secondary cardiac tumours (i.e., cardiac metastases) are more common (incidence of 1.2\%) than primary cardiac tumours. Metastatic cardiac tumours mostly originate from carcinoma of all types, lymphoma, melanoma and sarcoma\cite{sarjeant_cancer_2003,bussani_cardiac_2007,goldberg_tumors_2013}. The low prevalence of such rare pathology induces challenges in treatment optimization and complicates the definition of a treatment consensus\cite{stergioula_multimodality_2023}. Patients with cardiac tumours are treated mainly by surgical resection, which is eventually, if applicable, combined with adjuvant chemotherapy\cite{chan_primary_2023}. In some rare cases, orthotopic heart transplantation or cardiac autotransplantation were performed\cite{grandmougin_total_2001, chan_primary_2023}. 

Surgical interventions for the treatment of cardiac tumours are complex and carry procedural risks. As an alternative to surgery, cardiac tumours could potentially also be treated with a stereotactic body radiation therapy (SBRT)\cite{batumalai_mrlinac_2023}. SBRT is a noninvasive treatment approach in which a radiotherapy treatment procedure is conducted with high precision using higher radiation doses (with steep dose gradients) per fraction. The procedural risks of invasive interventions could therefore be avoided, while still treating cardiac tumours. The heart and adjacent mediastinal structures are highly sensitive to radiation, making precise targeting crucial to minimize toxicity. From thoracic radiotherapy, it is known that the heart is at risk to develop radiotherapy-induced cardiotoxicity\cite{yusuf_radiation-induced_2011, ratosa_cardiotoxicity_2019, zou_radiotherapy-induced_2019}. It is therefore essential to minimize the treatment target volume, dose to organs-at-risk (OAR) and attain a highly conformal dose distribution, which can be facilitated by accurately visualizing the anatomy (i.e., treatment target and OAR) prior to and during treatment using on-board imaging.
% It is therefore crucial to minimize the treatment target volume and attain a highly conformal dose distribution. It is therefore essential to accurately determine the position of the treatment target and visualize surrounding healthy organs-at-risk (OAR) using daily imaging. 

Medical imaging during a radiotherapy procedure is primarily driven by computed tomography (CT) during simulation and cone beam CT (CBCT) for daily adaptation. While CT imaging provides clear definition of air, bone and soft tissue, its limited soft tissue contrast makes target delineation challenging. Magnetic resonance imaging (MRI) provides superior soft-tissue contrast and facilitates the accurate visualization of the treatment target and OAR for a radiotherapy procedure\cite{chandarana_emerging_2018}.

The development of hybrid MRI and linear accelerator systems (MR-linac), such as the 1.5 T Unity\cite{lagendijk_mrilinac_2008,lagendijk_magnetic_2014} (Elekta AB, Stockholm, Sweden) and MRIdian\cite{mutic_viewray_2014} (ViewRay, Oakwood Village, OH), introduced on-board MR imaging that facilitates MR-guided radiotherapy (MRgRT). An MRgRT treatment approach could therefore facilitate highly conformal dose administration, while sparing surrounding tissue, due to the accurate visualisation of the daily anatomy using MRI and the ability to tailor the SBRT treatment upon to based on anatomy changes. Additionally, the MR-linac is also capable of mitigating physiological motion during treatment, such as respiratory motion\cite{grimbergen_gating_2023} and even, in a research configuration, cardiac motion\cite{akdag_first_2022,akdag_experimental_2024}.

There is some evidence that demonstrate the feasibility of MRgRT treatments of cardiac tumours. Patients with cardiac tumours have been treated with MRgRT using the 1.5 T Unity MR-linac\cite{pomp_sarcoma_2021,noyan_cardiac_2022,batumalai_mrlinac_2023,chen_case_2024} and 0.35 T MRIdian MR-linac\cite{gach_lessons_2019,corradini_mr-guided_2021,michalet_stereotactic_2024}. The amount of evidence is, however, limited due to the rarity of the clinical indication and the new strategy of using MR-guided SBRT for the treatment of patients with cardiac tumours. 
% MRI-guided radiotherapy (MRgRT) is a non-invasive treatment option offering direct target visibility that is rapidly emerging for patients with cardiac tumours to facilitate highly conformal dose administration, while sparing healthy surrounding tissue. The purpose of this study is to conduct a patterns-of-care analysis of cardiac MRgRT treatments through a survey of users of the 1.5T Unity MR-linac (Elekta AB, Sweden) that treated patients with cardiac tumours
In this study, we aimed to conduct a patterns of care (i.e., technical and workflow aspects) analysis of cardiac MRgRT treatments on the 1.5 T Unity MR-linac and consolidate international real world experience to inform future treatment protocols and standardization efforts on the MR-linac. This patterns of care analysis was conducted through a survey of worldwide users of the 1.5 T Unity MR-linac that treated patients with cardiac tumours. The adopted strategies for cardiac MRgRT treatments were characterized via this survey, which will contribute to treatment harmonization on the 1.5 T Unity MR-linac.

\section{Material and Methods}\label{sec:Methods}
To conduct our patterns of care survey, a questionnaire was developed inspired by the RATING-framework\cite{hansen_radiotherapy_2020} and the STOPSTORM.eu consortium survey\cite{grehn_stereotactic_2023}. The questionnaire was only distributed to centers that use the Elekta 1.5 T Unity MR-linac for the treatment of patients with primary or secondary cardiac tumours.
The questionnaire was distributed in October 2023 and data was collected until October 2024. The participants completed the survey sheets while complying with the ethical regulations of their institution.

The questionnaire included 30 questions divided over six sections, which can be found in the Supplementary Material. Close-ended and multiple choice questions were used to ensure harmony between given answers and simplify data processing for comparison purposes. The first section contained anonymous patient cohort data, including age, gender, disease type and indication for treatment. The remaining five sections considered technical details regarding diagnosis, radiotherapy treatment simulation, irradiation and treatment outcome. The collaborators were requested to score the on-board acquired MRI images (3D pre-treatment and cine images during treatment) according to a four point Likert-scale. The collaborators were also asked to provide feedback on their treatment workflow and utilized hard- and software applications/tools. Questions that remained unanswered after providing reminders were labeled as `Not Provided (N.P.)'.

\section{Results}\label{sec:Results}
A total of twelve questionnaires were completed covering twelve cardiac radiotherapy treatments on the 1.5 T Unity MR-linac between 2021--2024. Five of the reported twelve cases were previously described\cite{pomp_sarcoma_2021,noyan_cardiac_2022,batumalai_mrlinac_2023,chen_case_2024}. The survey sheets were collected from six international institutions: Australia (two institutions), United States of America (two institutions), The Netherlands (single institution) and Türkiye (single institution). The distributed survey sheet and all collected responses are provided as Supplementary data with this manuscript.

\subsection{Patient cohort and diagnostic imaging}
\begin{figure}[h]
    \centering
    \includegraphics[width=\linewidth]{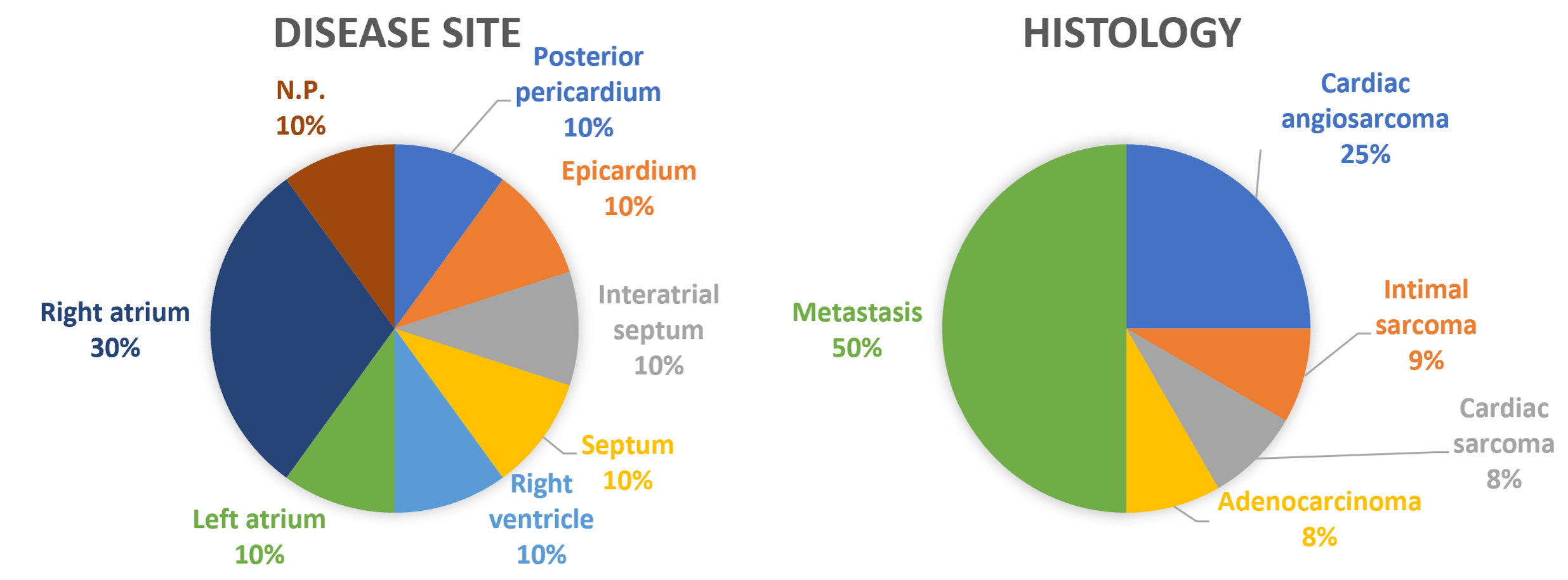}
    \caption{Overview of the disease sites and histology for the included treatment cases (N.P. = not provided).}
    \label{fig:estro-histology-diseasesite}
\end{figure}

This study included twelve patients (7 male, 5 female) with a median (range) age of 59 (16--81) years at the time of treatment. Four possible treatment indications could be reported: primary (50\%), in case of recurrence (50\%), post-operative (25\%) and palliative (42\%). The number of treatment indications ranged between one and three per patient (single targets only). Diagnostic modalities during the diagnosis phase included MRI (100\%), positron-emitted tomography (83\%), CT (67\%) and ultrasound (8\%). The reported distribution of sites and histology are shown in Figure \ref{fig:estro-histology-diseasesite}. The following metastatic lesions were reported: melanoma-derived metastasis, carcinoma - sarcomatoid differentiated, lung cancer, small cell lung cancer, adenocarcinoma, thymoma. For analysis, all metastatic lesions were grouped under a single 'Metastasis' entry.

\subsection{Treatment simulation}
\subsubsection*{Imaging for treatment planning}
CT and MRI simulators were used for treatment planning in all cases, with contrast-enhanced CT utilised in 27\% of the cases. In 55\% of all cases, respiratory-resolved 4D-CT was acquired (of which 33\% with a compression belt). Motion mitigation was omitted during all acquisitions on the MRI-simulator. In a single case, 2D cine-MRI was acquired for the investigation of respiratory motion. In 75\% of the cases, the images acquired on the CT-simulator were indicated as the planning master scan, while the MRI was indicated as the master scan in 8\% of the cases. The planning master scan was not provided in 17\% of the cases.

\subsubsection*{Treatment plan characteristics}
% GTV and PTV margin
The  gross target volumes (GTV) ranged from 4.1 to 186.6 cc with a median of 17.7 cc. An internal target volume (ITV) approach was employed in 55\% of the cases, with a compression belt used in 33\% of these cases. All reported clinical target volume (CTV) to PTV margins were isotropically defined ranging between 2--7 mm with a median CTV-to-PTV margin of 3 mm.

\begin{figure}[!h]
    \centering
    \includegraphics[width=\linewidth]{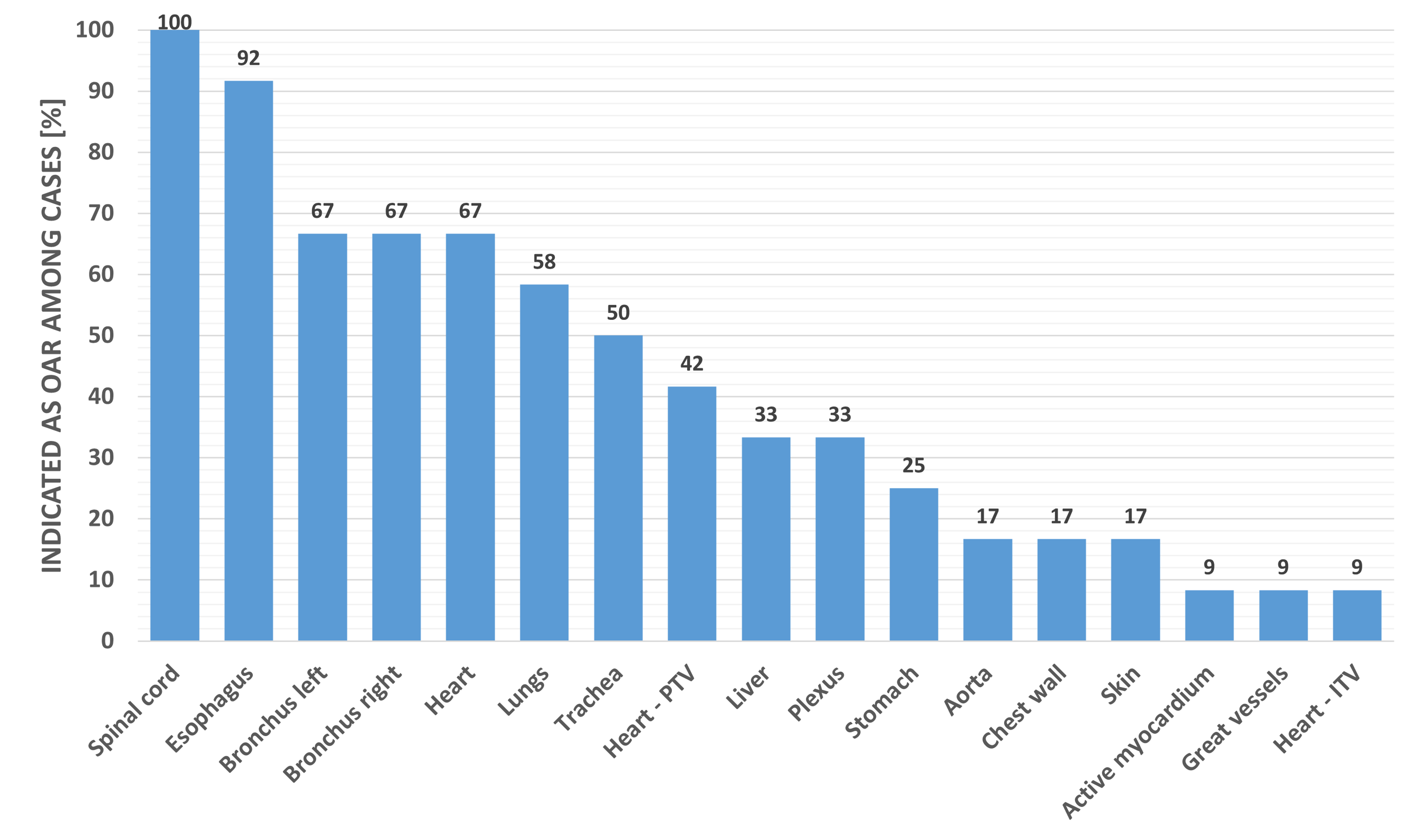}
    \caption{An overview of the indicated organs-at-risk (OAR) among the reported treatment cases. }
    \label{fig:OAR-bar-chart}
\end{figure}

% Organs at risk
Twelve distinct OAR were reported. The distribution of reported OARs is shown in Figure \ref{fig:OAR-bar-chart}. The reported dose constraints for all OARs are shown in Table S2. Two institutions, responsible for three treatment cases, reported that they conducted the delineation tasks using MIM (MIM Software Inc., Cleveland, OH, USA). The remaining institutes only reported the use of Monaco throughout the treatment planning procedure.

\begin{figure}[!ht]
    \centering
    \includegraphics[width=\linewidth]{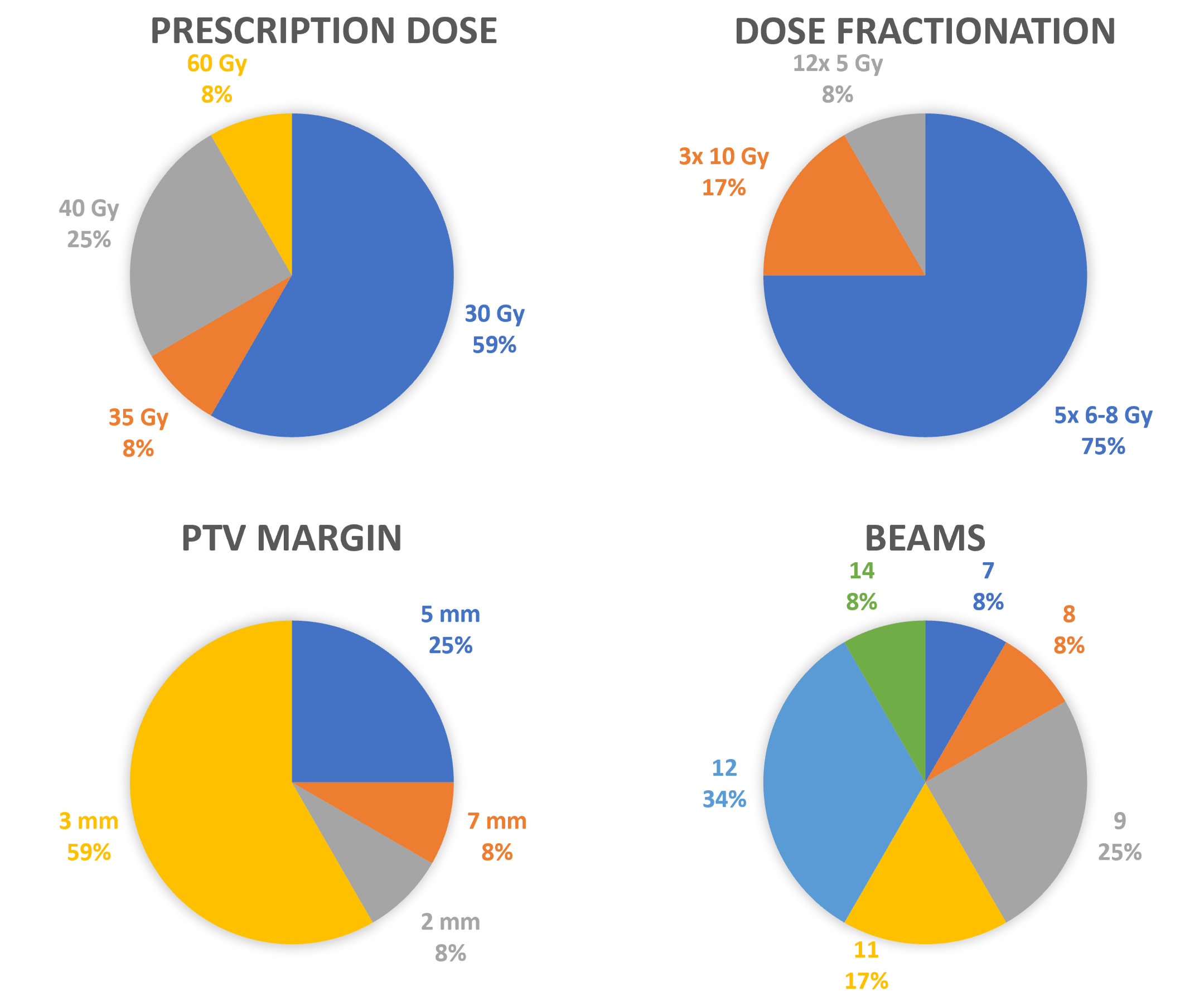}
    \caption{A summary of key treatment planning characteristics of the cardiac radiotherapy treatment planning procedure.}
    \label{fig:TP-PieCharts}
\end{figure}

% Dose calculation
The treatment plans were mainly calculated in Monaco v5.4 or higher. A calculation uncertainty of 1\% was used in 83\% of the plans, while 17\% were calculated with an uncertainty of 3\% per control point. The treatment plans were calculated with dose calculation grid sizes of 2 (17\%), 2.5 (33\%) and 3 (50\%) mm. The total prescribed dose ranged from 30 to 60 Gy. The delivered dose per fraction ranged between 5--10 Gy with a median dose per fraction of 6 Gy. The amount of segments ranged between 50--136 with a median of 82. An overview of the resulting treatment plan characteristics captured in this survey are represented in Figure \ref{fig:TP-PieCharts}.

% Treatment plan calculation/characteristics and acceptability
The target dose levels were not achieved in 17\% of the cases and the OAR dose levels were exceeded in 17\% of the cases. In 25\% of the treatment cases, the PTV was intentionally under-dosed for tissue-sparing purposes. The treatment plan for 25\% of the treatment cases were accepted with minor violations, while the remaining 75\% of the cases had no violations regarding plan acceptability.

% The median[range] dose per fraction was 6[5-10] Gy. The median [range] number of segments in the resulting treatment plans was 82[50-136]. The treatment plans were mainly calculated in Monaco v5.4 or higher with a dose calculation uncertainty of 1\% per plan, except for a single plan that was calculated with an uncertainty of 3\% per control point, on grid sizes between 2-3 mm. A single institute used MIM (MIM Software Inc., Cleveland, OH, USA) for contouring tasks. The target dose levels were not met in 18\% of the cases for organs-at-risk sparing purposes.

% \begin{figure}[!ht]
%     \centering
%     \includegraphics[width=0.70\linewidth]{Figure 1_07032024.pdf}
%     \caption{This figure shows a summary of key treatment planning characteristics of the cardiac radiotherapy treatment planning procedure.}
%     \label{fig:estro1}
% \end{figure}

\subsection{Radiotherapy treatment}
% Patient positioning
A day-zero fraction was conducted on the MR-linac in 75\% of the treatment cases. The patient was positioned using a vacuum mattress (often in combination with a wing board and wedge) in 45\% of the cases. In a single case, a positioning tattoo was used combined with a sagittal positioning laser. The patient’s arms were positioned up in 50\% of the cases. 

% Daily imaging
Daily T\textsubscript{1}-weighted and T\textsubscript{2}-weighted MRI was acquired for 25\% and 83\% of all cases during all fractions, respectively. A custom in-house optimized 3D T\textsubscript{2}-weighted MRI was used in a single treatment case. Additionally, the 3DVane image sequence was employed in 25\% of all cases during all fractions. Diffusion-weighted imaging was performed in one case during one fraction.

\begin{figure}[!ht]
    \centering
    \includegraphics[width=\linewidth]{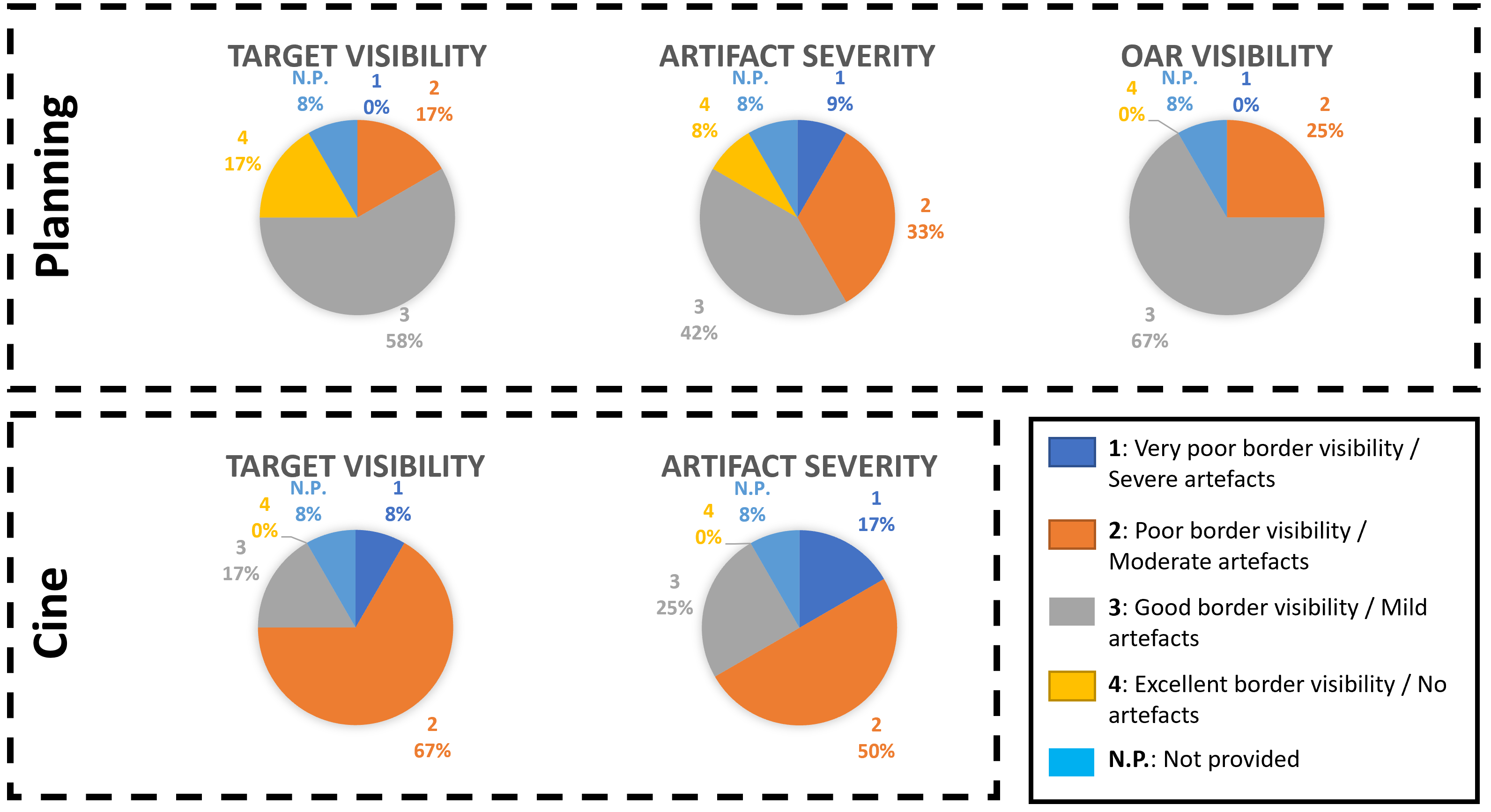}
    \caption{The visual scoring of the used daily and cine MRI images, which were acquired on the MR-linac, during the MRI-guided cardiac radiotherapy treatment based on a four-point Likert scale.}
    \label{fig:DailyMRScores}
\end{figure}

% Image quality scores
The image quality scores of the acquired planning and cine MR-images are shown in Figure \ref{fig:DailyMRScores}. Among planning images, 75\% were rated good or better for target visibility and 67\% for OAR visibility. Artifacts were mild or absent in 50\% of the images. Only 17\% of cine MR images provided good treatment target border visibility, while 67\% exhibited moderate to severe image artifacts.
% The acquired MR-images were scored by the responsible physician regarding the target visibility, OAR visibility and artifact severity based on a four-point Likert scale (Figure \ref{fig:DailyMRScores}).

% Daily adaptation
The adapt-to-position (ATP) approach was used in 16\% of the delivered fractions, while the remaining fractions (84\%) were adapted according to the adapt-to-shape (ATS) approach. Deformable image registration was used in all cases with the ATS approach to warp the reference to the daily anatomy. The CT scans were indicated as first reference for 58\% of the treatment cases, while the remaining 42\% of the cases used the MRI scans as first reference. In 33\% of the cases, both CT and MRI scans were used as reference for the remaining fractions. For 25\% of the cases, the MRI from previous fractions served as reference scan. In 8\% of the cases, the MRI from the first fraction was used as reference for the remaining fractions.

\begin{figure}[!ht]
    \centering
    \includegraphics[width=\linewidth]{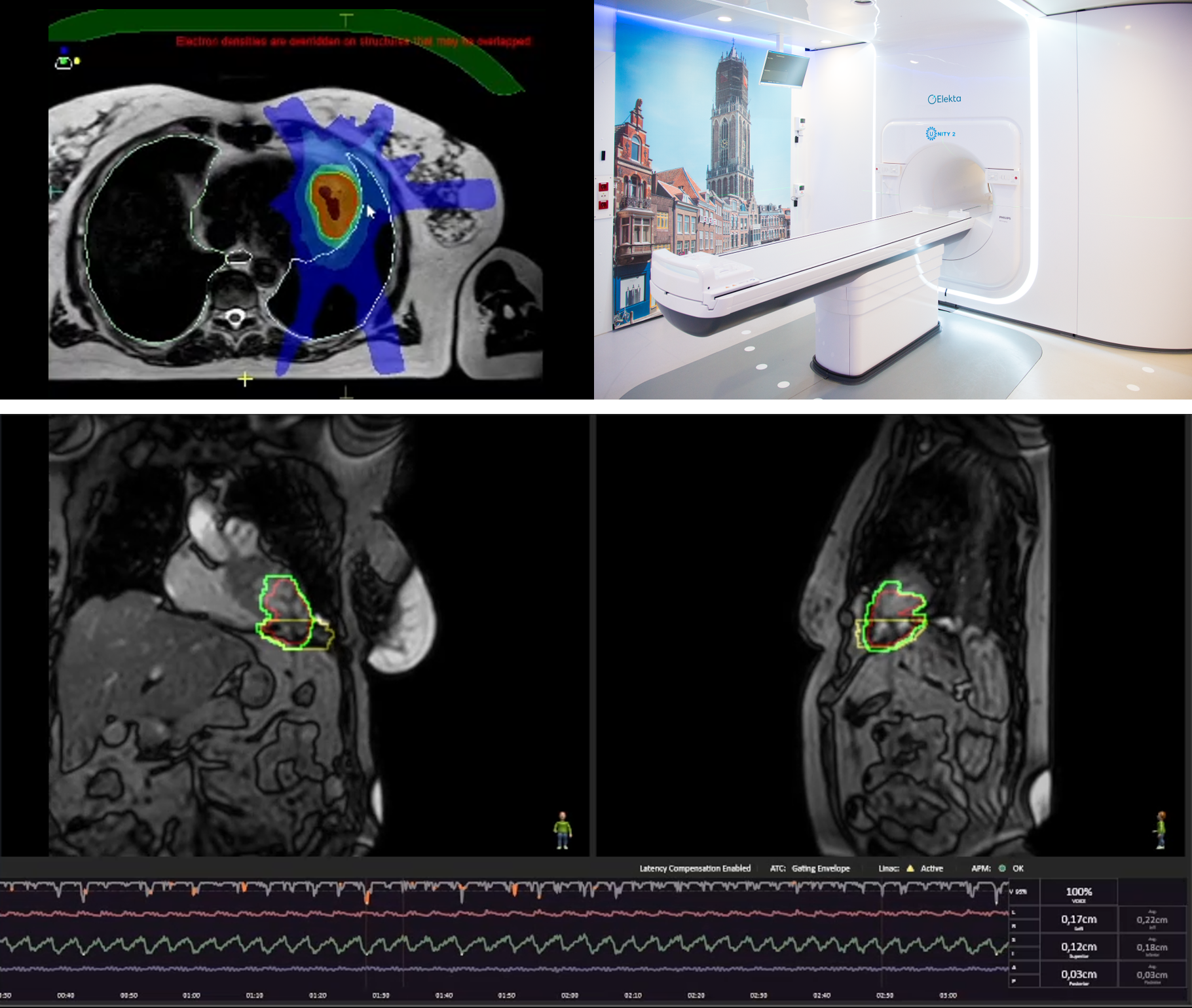}
    \caption{\textit{Top left}: The calculated treatment plan visualized in the axial plane within the Monaco treatment planning system. \textit{Top right}: A picture of the treatment room. \textit{Bottom row}: The sagittal and coronal cine images within the Comprehensive Motion Management workflow system. The gating envelope, gross target volume and gating surrogate are indicated with the green, red and yellow contours, respectively.}
    \label{fig:cmm}
\end{figure}

% Intrafraction 
During dose delivery, motion monitoring was conducted using cine MRI acquisitions in 92\% of the treatment cases. Vendor-provided motion monitoring software was used in 82\% of the cases with motion monitoring. Intrafractional respiratory motion was mitigated by radiation beam-gating for one of the patients using a novel vendor-provided gating solution for clinical routines. The newly introduced solution provides tools for intrafractional treatment adaptations, real-time motion monitoring and mitigation\cite{smith_commissioning_2025}. A surrogate structure encompassing the apex of the heart and its surroundings was delineated for improved treatment target tracking performance (Figure \ref{fig:cmm}).

The overall treatment times ranged between 36.3--60.0 minutes with a median overall treatment time of 45.2 minutes.

% Deformable image registrations of daily MRI with the master planning CT image dataset were performed during the first fraction. Daily replanning was performed with the adapt-to-shape(adapt-to-position) approach in 83\% (17\%) of the reported treatments (Figure \ref{fig:estro-daily-replanning-monitoring}). Cine-MRI motion monitoring was used in 91\% of the reported cases (80\% using a vendor-provided sequence). 

% \begin{figure}[!ht]
%     \centering
%     \includegraphics[width=\linewidth]{Fig5-CineScores.png}
%     \caption{The visual scoring of the used cine MRI images for motion monitoring during the MRI-guided cardiac radiotherapy treatment based on a four-point Likert scale.}
%     \label{fig:CineMRScores}
% \end{figure}

\subsection{Early patient outcome}
The collaborators were able to report the treatment outcomes of 50\% of the treatment cases. All, but one case, attained local control using MRI-guided cardiac radiotherapy. No serious adverse events ($\geq$ grade 3) were reported.

% As overall feedback, the collaborators mentioned the desire for improved cardiorespiratory motion compensation/mitigation during imaging and delivery. The desire to be able to use MRI acquisitions in the cardiac imaging planes (i.e., short axis, long axis and four chamber-view) in the treatment workflow was reported for improved target delineations. --> Moved to discussion

% Median[range] mean overall treatment times of 45.8[36.3-80] minutes were reported. 45\% of the collaborators reported treatment outcomes of their conducted treatments. All, but one, attained local control with MRI-guided cardiac radiotherapy. No serious adverse events (>grade 3) were reported. Better cardiorespiratory motion compensation during imaging and delivery, as well as MRI acquisitions along the cardiac axes during planning for improved target delineation were returned as suggestions for future cardiac MRgRT improvements.

\section{Discussion}\label{sec:Discussion}
% Introductory summary of conducted works
This study presents the first comprehensive survey of emerging patterns of care for cardiac MRgRT. Here, users from six different institutes shared their experience in treating patients with cardiac tumours on the 1.5 T Unity MR-linac. The shared experiences show that MRI-guided cardiac radiotherapy is technically feasible for oncological indications and thus has the potential to resolve the practical limitations and be a safe alternative of invasive treatment approaches. 

% Most common dose prescription
The most treatments were delivered in five fractions with prescription doses ranging between 30--40 Gy. The close agreement of the aforementioned dose prescriptions within this survey can therefore be considered as a base for a dose prescription consensus, while there are no international guidelines (yet) for MR-guided radiotherapy treatments of cardiac tumours.

% The effect of cardiorespiratory motion during simulation
The cardiac treatment targets are subject to cardiorespiratory motion, and thus prone to motion-induced dosimetric uncertainties. Cardiac and/or respiratory motion can be addressed by applying motion mitigation strategies during treatment simulation and/or dose delivery. In the treatment simulation phase, CT-simulator and MRI-simulator were used in all patients. However, only 55\% of the acquired CTs were respiratory-resolved. No motion mitigation approaches were applied during MR image acquisitions. Cardiac motion was not considered in the preparation phase, which might be due to the absence of the appropriate tools to facilitate cardiac motion mitigation during treatment simulation.

% The benefit of on-board MRI
The on-board MRI-system permitted the direct visualization of the treatment target and surrounding anatomy with great soft tissue-contrast. One of the collaborating institutes had the ability to mitigate intrafractional motion with the newly available vendor-provided gating solution in a single treatment case, which demonstrated that it is technically feasible to mitigate respiratory motion during dose delivery for the treatment of cardiac tumours. A gating surrogate was delineated comprising the apex of the heart for the estimation of respiratory motion, which was challenging due to its complex anatomy. 

% The quality of current available image sequences
The reported visibility and artifact severity scores have shown that the current image quality for daily imaging is sufficient, while the majority of the cine images were scored worse. The scores for the acquired images indicated that improvements of the imaging sequences are desired for cardiac MRgRT.

% Image corruption by presence of implants and/or CIEDs
One patient had a bioprostethic mitral valve that induced imaging artifacts as it locally disrupts the homogeneity of the magnetic field. The presence of the implant required a tailored treatment, which was done within a multi-disciplinary collaboration. Patients, especially with cardiac arrhythmias, potentially carry a cardiac implantable electronic device (CIED), which are generally distally placed of the clavicle. These devices, with the attached leads, can induce larger imaging artifacts and require tailored approaches for imaging\cite{vigen_recommendations_2021} and treatment planning\cite{miften_management_2019} purposes according to certain safety guidelines\cite{keesman_multi-institutional_2024}.

% The indication of OARs
A total of twelve distinct OARs were reported in this patterns of care analysis. The spinal cord was indicated by all collaborators, while 92\% of the collaborators indicated the esophagus as OAR. The high occurrence of these OARs provides a base for consensus that these two OARs should always be indicated during MRgRT treatment of cardiac tumours. 

The heart was indicated as OAR in four different fashions: the whole heart, whole heart excluding the PTV, whole heart excluding the ITV, or only the active myocardium. The variability between collaborators regarding the definition of the heart as OAR suggests that there is no consensus, which warrants further discussions and research.
In this study, we were able to pool the available data using a questionnaire and assess the technical feasibility of MRgRT for the treatment of cardiac tumours. There are however some aspects in this patterns of care analysis that made a thorough analysis challenging.

% Low sample size and restricted treatment outcome analysis
The rare indication of cardiac tumours and the novelty of treating cardiac tumours with MRgRT limited the amount of data and we were only able to assess the technical feasibility using data of twelve treatment cases, which therefore constrains the generalizability of the findings. Additionally, due to ethical restrictions within collaborating institutes, 50\% of collaborating institutes were not able to share results of their follow-up. The ability to draw robust conclusions about clinical effectiveness or toxicity was therefore limited.

% Feedback from collaborators
The collaborators were requested to provide feedback on the use of the 1.5 T MR-linac for the treatment of cardiac tumours. As overall feedback, the collaborators mentioned the desire for improved cardiorespiratory motion compensation/mitigation during imaging and delivery. The desire to be able to use MRI acquisitions in the cardiac imaging planes (i.e., short axis, long axis and four chamber-view) in the treatment workflow was also reported. The rationale is that MRI acquisitions in the cardiac imaging planes can facilitate more accurate target delineations. Novel vendor-provided image acquisition and processing solutions would be required to facilitate cardiac MRI during a cardiac MRgRT treatment workflow.

% Comparison with other works - VT focused
This study focused on MRgRT treatment of cardiac tumours using the 1.5 T MR-linac. Treatment cases have been reported in literature in which patients were treated with radiotherapy on either a conventional CT-linac\cite{fields_treating_2017,krishnan_cardiac_2020} or 0.35 T MR-linac (ViewRay, Oakwood Village, OH)\cite{gach_lessons_2019,sim_mr-guided_2020,corradini_mr-guided_2021,michalet_stereotactic_2024} for the treatment of cardiac tumours. More recently, seven patients with cardiac tumours were treated with MRgRT in a single institution using the 1.5 T MR-linac. These seven patients with cardiac tumours were treated according to a split-course hypofractionated RT with concurrent chemotherapy with promising clinical outcomes\cite{zheng_safety_2025}. Patients with ventricular tachycardia (VT) were also treated on a 0.35 T MR-linac by delivering a single high-dose fraction of 25 Gy to the treatment target\cite{mayinger_first_2020,bianchi_magnetic_2024}. All cardiac MRgRT treatments on the 0.35 T MR-linac were done with intrafraction respiratory motion management. All VT patients carried a CIED and no related adverse events were observed.

% Concluding paragraph
Despite the study limitations, a first global multi-institutional survey exploring the feasibility and emerging practices of cardiac MRgRT using a 1.5 T MR-linac was presented. A comprehensive overview of current clinical approaches was offered, from imaging and simulation to treatment delivery and adaptation. The inclusion of international centers enhances its relevance and provides a foundation for cross-institutional harmonization. By identifying common practices and areas of divergence, this study offers valuable baseline data that can inform future consensus guidelines and protocol development for MRgRT in rare cardiac oncological indications, including a possible extension to the treatment of cardiac arrhythmias. The development of treatment recommendations could be supported by conducting a thorough patient-outcome analysis that encompasses a longer period of time.

% This study provides a knowledge baseline for future improvements of cardiac radioablation treatments on the 1.5 T MR-linac. 

%% Use \subsubsection, \paragraph, \subparagraph commands to 
%% start 3rd, 4th and 5th level sections.


%%       \cite{<label>} ==> [1]

%%

\clearpage
\bibliographystyle{model3-num-names} 
\bibliography{references}
\biboptions{sort&compress}



\end{document}